\documentclass[seceq,preprint]{ptptex}

\notypesetlogo  %comment in if to eliminate PTPTeX logo
\preprintnumber[3cm]{KYUSHU-HET-109}

\title{
Charged Lepton and Down-Type Quark Masses in 
$SU(1,1)$ Model and the Structure of Higgs Sector
}

\author{
Kenzo INOUE \footnote{E-mail: inou1scp@mbox.nc.kyushu-u.ac.jp}
and Naoki YAMATSU\footnote{E-mail: yamatsu@higgs.phys.kyushu-u.ac.jp}
}

\inst{
Department of Physics, Kyushu University, Fukuoka 812-8581, Japan
}

\abst{
The simplest noncompact group $SU(1,1)$,
when introduced as a symmetry group of the generations 
of quarks and leptons in the framework of a vector-like theory, 
gives an excellent viewpoint on low energy physics.
The minimal setup of the scheme, however,
 gives phenomenologically unacceptable prediction on the Yukawa coupling
 matrices.
This suggests the higgs sector has richer structure
than we expect from the success of MSSM.
The natural extension of the scheme, which has doubled structure
in the higgs sector, is formulated.
The framework admits this
 extension in a restrictive way.
The possible patterns of Yukawa couplings are classified
and the expressions of the coupling matrices are presented.
}

\begin{document}
\maketitle

\section{Introduction}

The most mysterious nature in low energy physics is a simple repetition
of the three generations of quarks and leptons.
Nonetheless, they have a remarkable mass structure, 
the inter-generation mass hierarchy.
Such a subtlety should have its origin in a definite principle in Nature.
Thus, it will be legitimate to search for the principle
based on the symmetry which governs the generations,
that is, horizontal symmetry $G_H$.
\cite{ref:horizontal_symmetry,ref:Frog-Nielsen}

One of the possible models was proposed based on the noncompact gauge
symmetry $G_H=SU(1,1)$. \cite{ref:su11-1} 
This model is a vector-like \cite{ref:left-right_symmetry} 
realization \cite{ref:KAZUO} 
of the minimal supersymmetric standard model (MSSM). 
\cite{ref:MSSM,ref:MSSM-review}
The model contains minimal numbers of vector-like matter 
multiplets $F$ and $\bar F$, 
which belong to
infinite dimensional unitary representations of $SU(1,1)$.
The appearance of three chiral generations results from the
spontaneous breakdown of $SU(1,1)$. 
What is more, this symmetry breaking naturally 
realizes the hierarchy in the Yukawa couplings of  higgses 
to quarks and leptons 
based on the group theoretical structure of $SU(1,1)$.

Since the original $SU(1,1)$ model was proposed, 
some attempts have been made 
 \cite{ref:su11-2,ref:su11-3}
to examine the phenomenological feasibility of the model, 
and also to inquire into the further potentialities of the model.
Through these analyses, it was shown that
this model reproduces, even under the simplest choice of the parameters, 
the quark and lepton mass hierarchies at least
qualitatively.
But quantitatively, it was not able to exactly do so for 
down-type quarks and charged leptons simultaneously, because their 
observed mass structures are somewhat different and 
the simplest choice of the parameters cannot produce this difference. 
One may imagine the relaxation of the restriction on the parameters
solves this difficulty.
We have examined it and found
that the minimal setup of the model gives definite prediction
on the mass ratios of charged leptons and down-type quarks
which is phenomenologically unacceptable
for any value of the parameters. 

In this paper, we first clarify the reason why the 
 minimal $SU(1,1)$ model does not reproduce both of the 
mass hierarchies.
As we will see,
if we introduce, according to the success of MSSM,\cite{ref:MSSM-success}
only one down-type higgs multiplet,
the structure of its Yukawa couplings to
down-type quarks and charged leptons is essentially controlled
by the $SU(1,1)$ group property, which turns out to be too rigid
to compromise by the remaining freedom of the model.
This suggests that
the acceptable model should have at least doubled higgs structure.
In order to reproduce MSSM at low energy, they must mix each other,
and realize only one combination as a chiral down-type higgs.
The main purpose of this paper is to examine the possible extension 
of the higgs sector 
within the basic framework of the model
to have
in some extent the flexibility for the structure of Yukawa couplings.

In section 2, we give a setup of the minimal $SU(1,1)$ model 
for the fixing of the notation. 
In section 3, we discuss a property of the minimal 
model and show the reason why this
model fails.
In section 4, we examine the extension of the higgs sector
and formulate the novel mixing scheme.
We classify the possible patterns of mixing, which are related to
the allowed patterns of the Yukawa couplings.
The resulting structures of the Yukawa couplings are given in section 5.
Section 6 is devoted to the summary and discussions.

\section{SU(1,1) model}

\begin{table}[t]
\begin{center}
   \begin{tabular}{c||c|c|c|c|c|c|c||c}  
     &  $Q$   & $\bar{U}$ & $\bar{D}$ & $L$ & 
           $\bar{E}$ & $H$ & $H'$ &    $\Psi$'s\\ \hline \hline
     $SU(1,1)$ &  $+\alpha$ & $+\beta$ & $+\gamma$ & $+\eta$ & 
           $+\lambda$ & $-\rho$ & $-\sigma$ & finite  \\ \hline
     $SU_3\times SU_2$& $(3,2)$ & $(3^*,1)$ & $(3^*,1)$ & $(1,2)$ & 
           $(1,1)$ & $(1,2)$ & $(1,2)$ &  $(1,1)$ \\ \hline
     $Y/2$ &  $+1/6$    & $-2/3$ & $1/3$ & $-1/2$ & $+1$ & $+1/2$ &
    $-1/2$ &  $0$ \\ 
   \end{tabular}
\end{center}
\caption{The $SU_3\times SU_2\times U_1 \times SU(1,1)$ assignment for the 
multiplets in the $SU(1,1)$ model}
\label{tb:weight}
\end{table}
Let $Q_\alpha$ be a multiplet carrying the 
$SU_3\times SU_2\times U_1$ quantum numbers
of quark doublet $q$ and
belonging to an infinite-dimensional
unitary representation of $SU(1,1)$ with a positive lowest weight $\alpha$.
Its conjugate $\bar Q_{-\alpha}$ has a
 negative highest weight $-\alpha$;
\begin{equation}
Q_\alpha=\{q_\alpha,q_{\alpha+1},\cdots\} ,\hspace{2em}
\bar Q_{-\alpha}=\{\bar q_{-\alpha},\bar q_{-\alpha-1},\cdots\} .
\end{equation}
The MSSM superfields $q,$ $\bar{u},$ $\bar{d},$ $\ell,$
$\bar{e},$ $h,$ $h'$ are embedded into
\begin{equation}
Q_\alpha,\ \bar{U}_\beta,\ \bar{D}_\gamma,\ L_\eta,\ \bar{E}_\lambda,
\ H_{-\rho},\ {H}_{-\sigma}' ,
\label{Q}
\end{equation}
respectively.
For the vector-like nature of the model, 
we also have the conjugate multiplets
\begin{equation}
\bar{Q}_{-\alpha},\ U_{-\beta},\ D_{-\gamma},\ \bar{L}_{-\eta},\
 E_{-\lambda},\ \bar{H}_{\rho},\ \bar{H}'_{\sigma} .
\label{barQ}
\end{equation}
In addition to (\ref{Q}) and (\ref{barQ}), 
we need some set of finite-dimensional non-unitary 
multiplets $\Psi$'s of a type
\begin{equation}
\Psi=\{\psi_{-S},\psi_{-S+1},\cdots,\psi_{S-1},\psi_S\}.
\end{equation}
They are $SU_3\times SU_2\times U_1$ singlets, and are responsible to the
spontaneous breakdown of $SU(1,1)$.
They are indispensable to realize the chiral world at
low energy from originally vector-like theory.
For example, the coupling $Q\bar Q\Psi^F$ with
the vacuum expectation value (VEV) $\langle\psi^F_{-3}\rangle$
of $\Psi^F$
generates three generations of chiral quark doublets
$q_m\equiv q_{\alpha+m}\ (m=0,1,2)$, 
because $q_m$'s disappear from the mass operator 
$Q\bar Q\langle\Psi^F\rangle$ owing to the weight conservation.
The other components of
$Q$ form with the components of $\bar Q$
the huge 
Dirac mass terms $\sum_{r=0}^\infty M_r\hspace{.5mm} q_{\alpha+3+r}
\hspace{.5mm}\bar q_{-\alpha-r}$
with mass $M_r$ blowing up in the limit $r\rightarrow\infty$ as 
$M_r\propto\langle\psi^F_{-3}\rangle\hspace{.5mm}r^{S^F}$ 
by the highest weight $S^F$ of $\Psi^F$.\cite{ref:su11-1}
Therefore, the superpotential
\begin{equation}
W_0=Q\bar{Q}\Psi^F+\bar{U}U\Psi^F+\bar{D}D\Psi^F
+L\bar{L}\Psi^F+\bar{E}E\Psi^F
+H\bar{H}(\Psi^{0}+\Psi^{1})+H'\bar{H}'(\Psi^{\prime 0}+\Psi^{\prime 1})
\label{W0}
\end{equation}
generates three chiral generations of  $q,$ $\bar{u},$ $\bar{d},$ $\ell,$
$\bar{e}$ through the VEV $\langle\psi^F_{-3}\rangle$.
The VEVs 
$\langle\psi^{0}_{0}\rangle$,$\langle\psi^{1}_{1}\rangle$
of $\Psi^{0}$,$\Psi^{1}$ and
$\langle\psi^{\prime 0}_{0}\rangle$,$\langle\psi^{\prime 1}_{1}\rangle$ 
of $\Psi^{\prime 0}$,$\Psi^{\prime 1}$
generate one generation of chiral higgs doublets $h$ and $h'$ as linear
combinations of infinite components of $H$ and $H'$,
respectively.
We assume all VEVs $\langle\psi\rangle$'s are roughly of order
$M\simeq 10^{16}$GeV to reproduce MSSM.

In this paper, 
we use the phase convention of the $SU(1,1)$ multiplets 
$F_\alpha$, $\bar G_{-\alpha}$ and $\Psi$
so that
the bilinears
\begin{equation}
\sum_{n=0}^\infty (-1)^n f_{\alpha+n}\bar g_{-\alpha-n} ,\ \ 
\sum_{n=0}^{2S}(-1)^n\psi^*_{-S+n}\psi_{-S+n} ,\ \ 
\sum_{n=0}^{2S}(-1)^n\psi_{S-n}\psi_{-S+n}
\end{equation}
are $SU(1,1)$ invariants.
The $SU_3\times SU_2\times U_1\times SU(1,1)$ assignment for
the multiplets is shown in Table \ref{tb:weight}.

The most general cubic superpotential
of the multiplets in (\ref{Q}) and (\ref{barQ}) that is
compatible with the 
$SU_3\times SU_2\times U_1\times SU(1,1)$ invariance is
\begin{eqnarray}
&&W_1=\bar{U}QH+\bar{D}QH'+\bar{E}LH'+U\bar{Q}\bar{H}+D\bar{Q}\bar{H}'
+E\bar{L}\bar{H}'\nonumber\\
&&\hspace{2.5em}+QQD+Q\bar{U}\bar{L}+\bar{Q}\bar{Q}\bar{D}+\bar{Q}UL ,
\label{W1}
\end{eqnarray}
where we have abbreviated the coupling constant of each operator.
The first line of (\ref{W1}) consists of the ordinary Yukawa
couplings and their ``mirror couplings''.
The second line contains operators which violate baryon-number and
lepton-number conservations.
Their low energy effects, however, are suppressed by the 
huge mass $M$.
The dangerous dimension-4 
($QL\bar D,\bar D\bar D\bar U,LL\bar E$) and dimension-5
($\bar U\bar U\bar D\bar E$)
operators that embarrass  MSSM \cite{ref:B-violation} 
are all forbidden by 
the $SU(1,1)$ symmetry because all of them have positive weights.
The coupling $\bar EH'H'$ is incompatible with $\bar ELH'$ in the 
weight constraint given below.

The outstanding property of the scheme is that all coupling constants
in (\ref{W0}) and (\ref{W1})
  can be taken to be real under suitable
phase convention of each multiplet.
This allows us to settle the invariance under space-inversion (P), 
charge-conjugation (C) and time-reversal (T) as a ``fundamental principle''
of Nature.
All of their violations observed in low energy physics are
attributed to the spontaneous breakdown of $SU(1,1)$.

The $SU(1,1)$ invariance gives a rigid constraint to the couplings
in (\ref{W1}).
For example, the coupling $\bar{U}QH$ is allowed only when the 
weights of each multiplet satisfy 
$\Delta\equiv\rho-\alpha-\beta = [\mbox{non-negative integer}]$.
It has been shown \cite{ref:su11-1}
 that the simple lowest coupling ($\Delta=0$) 
is indispensable for the operators in the first line of (\ref{W1})
to realize a  Yukawa coupling hierarchy.
This gives the restrictions
\begin{equation}
\alpha+\beta=\rho ,\hspace{2em}
\alpha+\gamma=\eta+\lambda=\sigma .
\label{weight-restriction}
\end{equation}
The low energy manifestation of the minimal model is
the MSSM multiplets
$q_m$, $\bar{u}_m$, $\bar{d}_m$, $\ell_m$, $\bar{e}_m$ $(m=0,1,2)$, $h$, $h'$,
that couple through the
effective superpotential
\begin{equation}
W_{\rm eff}=\sum_{m,n=0}^2\left(
y_u^{mn}\bar{u}_mq_nh+y_d^{mn}\bar{d}_mq_nh'
+y_e^{mn}\bar{e}_m\ell_nh'\right).
\end{equation}

We will not discuss in this paper on the mass structure of the 
up-type quarks
related to the up-type higgs $h$,
that has been nicely reproduced within 
the minimal setup of the model.\cite{ref:su11-2}
For the neutrino masses,\cite{ref:neutrino_review1,ref:neutrino_review2}
which are also related to $h$, 
the adequate extension of the scheme that generates the 
effective operator $\kappa_\nu^{mn}\ell_m h\ell_n h$ has been
realized. \cite{ref:su11-3} 
For a reliable discussion of the CKM matrix  \cite{ref:CKM}
and the MNS matrix,\cite{ref:MNS} we need to clarify
the both structures of up-type and down-type higgs sectors.
Thus, we concentrate, in this paper, on the down-type higgs sector,
which has been much problematic.

\section{Aspect of minimal model}

Let us start by giving the embedding of the MSSM chiral
multiplets into the SU(1,1) multiplets;
\begin{eqnarray}
q_{\alpha+i}=
\sum_{m=0}^2 q_mU_{mi}^q+[\mbox{massive modes}],&~~&
\bar{u}_{\beta+i}=
\sum_{m=0}^2 \bar{u}_mU_{mi}^u+[\mbox{massive modes}],\nonumber\\
\bar{d}_{\beta+i}=
\sum_{m=0}^2 \bar{d}_mU_{mi}^d+[\mbox{massive modes}],&~~&\nonumber\\
\ell_{\eta+i}=
\sum_{m=0}^2 \ell_mU_{mi}^\ell+[\mbox{massive modes}],&~~&
\bar{e}_{\lambda+i}=
\sum_{m=0}^2 \bar{e}_mU_{mi}^e+[\mbox{massive modes}],\nonumber\\
h_{-\rho-i}=hU_i+[\mbox{massive modes}],\hspace{2.4em} &~~&
h_{-\sigma-i}'=h'U_i'+[\mbox{massive modes}].
\end{eqnarray}
Since $Q,\bar{U},\bar{D},L,\bar{E},H,H'$ are unitary representations of
SU(1,1), all coefficients $U_i$'s should be the row vectors in the
unitary matrices. Thus, they satisfy 
\begin{equation}
\sum_{i=0}^\infty U_{mi}^{q*}U_{ni}^q=\delta_{mn},\ \ \ \mbox{etc}.
\end{equation}
The Yukawa coupling of down-type higgs $h'$ to leptons is derived by
extracting the massless modes from the $SU(1,1)$ invariant coupling 
\begin{equation}
y_E\bar{E}LH'=
y_E\sum_{i,j=0}^\infty
C_{i,j}^E\bar{e}_{\lambda+i}\ell_{\eta+j}h'_{-\sigma-i-j}
\to\sum_{m,n=0}^2y_e^{mn}\bar{e}_m\ell_nh' ,
\end{equation}
where $C_{i,j}^E$ is the Clebsch-Gordan(C-G) coefficient.
Therefore, the coupling matrix $y_e^{mn}$ has a general form
\begin{equation}
y_e^{mn}=y_E\sum_{i,j=0}^\infty C_{i,j}^EU_{mi}^eU_{nj}^\ell U_{i+j}',
\hspace{2em}m,n=0,1,2.
\label{y-general}
\end{equation}
For the C-G coefficient $C_{i,j}^E$, 
we give here, for the later convenience, 
the general expression which covers the non-lowest coupling
($\Delta\geq 0$)
\begin{equation}
\bar E_{\lambda}L_{\eta}H'_{-\sigma-\Delta}
=\sum_{i,j=0}^\infty C^{\lambda,\eta}_{i,j}(\Delta)\ 
\bar e_{\lambda+i}\ell_{\eta+j}h'_{-(\sigma+\Delta)-i-j+\Delta} ,
\end{equation}
where $\sigma=\lambda+\eta$.
We note $C_{i,j}^{\lambda,\eta}(\Delta)=0$ for $i+j<\Delta$.
When $i+j\ge \Delta$, it is given by
\begin{eqnarray}
C_{i,j}^{\lambda,\eta}(\Delta)&=& (-1)^{i+j}
\sqrt{\frac{i!j!\Gamma(2\lambda+i)\Gamma(2\eta+j)}
{(i+j-\Delta)!\Gamma(2\sigma+i+j+\Delta)}} \nonumber  \\
&&\hspace{-0.5em}\times \sum_{r=0}^\Delta (-1)^r
\frac{(i+j-\Delta)!\Gamma(2\lambda)\Gamma(2\eta)}
{(i-r)!(j+r-\Delta)!r!(\Delta-r)!\Gamma(2\lambda+r)\Gamma(2\eta-r+\Delta)}
,\nonumber\\
\label{C-G-1}
\end{eqnarray}
which satisfies the symmetry relation
$C_{i,j}^{\lambda,\eta}(\Delta)=(-1)^\Delta C_{j,i}^{\eta,\lambda}(\Delta)$.
Thus, we have 
\begin{equation}
C_{i,j}^E=C_{i,j}^{\lambda,\eta}(0)=
(-1)^{i+j}\sqrt{\frac{(i+j)!\Gamma(2\lambda+i)\Gamma(2\eta+j)}
{i!j!\Gamma(2\sigma+i+j)}}.
\label{C^E}
\end{equation}

It is instructive to examine
a trivial case in (\ref{y-general})
where the higgs doublet $h'$ is
realized as a pure $i=0$ component of $h'_{-\sigma-i}$, that is,
$U_i'=\delta_{i0}$. 
This gives
\begin{equation}
y_e^{mn}=y_EC_{0,0}^EU_{m0}^eU_{n0}^\ell.
\end{equation}
Since the rank of this matrix is 1, only one generation of leptons,
expressed as $\ell=\sum_{n=0}^2\ell_nU_{n0}^\ell$ and
$\bar e=\sum_{m=0}^2\bar{e}_mU_{m0}^e$, has Yukawa coupling to $h'$, and
remaining two orthogonal generations decouple from $h'$.
Therefore, it is indispensable to
introduce the mixing in the realization of $h'$. 

Our basic ansatz is to give the mixing coefficients $U_i'$ of $h'$
the ``hierarchical'' structure of the form
\begin{equation}
U_i'=U_0'(-\epsilon')^ib_i(\sigma),
\hspace{2em}\epsilon'\lesssim 1 .
\label{eq:higgs_hierarchy}
\end{equation}
For quarks and leptons, we assume, for a while,
three generations are realized through 
the VEV $\langle\psi^F_{-3}\rangle$ of $\Psi^F$.
This gives
\begin{equation}
   U_{mi}^\ell = U_{mi}^{e} = \delta_{mi},\hspace{2em}
m=0,1,2 .
\label{eq:quark_lepton_no_hierarchy}
\end{equation}
The coupling matrix $y_e^{mn}$ is then given by
\begin{equation}
y_e^{mn}= y_EU'_0\epsilon'^{m+n}b_{m+n}(\sigma)
\sqrt{\frac{(m+n)!\Gamma(2\lambda+m)\Gamma(2\eta+n)}
{m!n!\Gamma(2\sigma+m+n)}}.
\label{y-e}
\end{equation}

Let us show how the mixing coefficients $U'_i$ are determined
from the couplings 
$H'\bar{H}'(\Psi^{\prime 0}+\Psi^{\prime 1})$ in (\ref{W0}). 
The VEVs
$\langle\psi_0'\rangle\equiv\langle\psi_0^{\prime 0}\rangle$ 
and $\langle\psi_1'\rangle\equiv\langle\psi_1^{\prime 1}\rangle$
produce the higgs mass
operators 
\begin{equation}
H'\bar{H}'\langle\Psi'^0\rangle+H'\bar{H}'\langle\Psi'^1\rangle
=\sum_{i=0}^\infty
\left(C_i^{(0)}\langle\psi_0'\rangle h'_{-\sigma-i}\bar{h}'_{\sigma+i}
+C_i^{(1)}\langle\psi_1'\rangle h'_{-\sigma-i-1}\bar{h}'_{\sigma+i}\right),
\label{H'mass}
\end{equation}
where $C^{(0)}_i$ and $C^{(1)}_i$ are the C-G coefficients.
Since the massless mode $h'$ is embedded in $H'$ in the form
$h'_{-\sigma-i}=U_i'h'+[\mbox{massive modes}]$,
the orthogonality of $h'$ to massive modes 
(the disappearance of $h'$ from the mass operators (\ref{H'mass}))
requires
the coefficients $U_i'$ to satisfy the recursion equation
\begin{equation}
\langle\psi_0'\rangle C_i^{(0)}U_i'+
\langle\psi_1'\rangle C_i^{(1)}U_{i+1}'=0.
\end{equation}
This equation  precisely realizes the ansatz (\ref{eq:higgs_hierarchy}) 
of the mixing coefficients 
with
\begin{equation}
\epsilon'=\frac{\langle\psi'_0\rangle}{\langle\psi'_1\rangle} ,
\hspace{2em}
b_i(\sigma)=\prod_{r=0}^{i-1}\frac{C_r^{(0)}}{C_r^{(1)}} .
\label{epsilon}
\end{equation}
For the relevant C-G coefficients,
we first give the general formula for the coupling 
\begin{equation}
H'_{-\sigma}\bar K'_\zeta\Psi'=\sum_{i,j=0}^\infty
D^{\sigma,\zeta}_{i,j}(S)\ h'_{-\sigma-i}\bar k'_{\zeta+j}
\psi'_{i-j+\bar\Delta} ,
\hspace{2em}\bar\Delta\equiv\sigma-\zeta ,
\end{equation}
which is allowed when the highest weight $S$ of $\Psi'$ satisfies
$S\ge |\bar\Delta|$.
For $-S\le i-j+\bar\Delta\le S$, we have
\begin{eqnarray}
D^{\sigma,\zeta}_{i,j}(S)&=&(-1)^j
\sqrt{\frac{i!j!(i-j+S+\bar\Delta)!(-i+j+S-\bar\Delta)!}
{\Gamma(2\sigma+i)\Gamma(2\zeta+j)}}\nonumber\\
&&\times\sum_{r=0}^{S-\bar\Delta}\frac{\Gamma(2\sigma+i+r)}
{(S-\bar\Delta-r)!(-i+j+S-\bar\Delta-r)!r!
(i-j+2\bar\Delta+r)!(i-S+\bar\Delta+r)!}~,\nonumber\\
\label{C-G-2}
\end{eqnarray}
and otherwise 
$D^{\sigma,\zeta}_{i,j}(S)=0$.
Thus, $C_r^{(0)}$ and $C_r^{(1)}$ are given by
\begin{eqnarray}
C_r^{(0)}=D_{r,r}^{\sigma,\sigma}(S_0) ,
\hspace{10mm}
C_r^{(1)}=D_{r+1,r}^{\sigma,\sigma}(S_1) ,
\end{eqnarray}
where $S_0$ and $S_1$ are the highest weights of $\Psi'^0$ and $\Psi'^1$.
For the lower values of $S$, they are
\begin{equation}
D_{r,r}^{\sigma,\sigma}(0)=(-1)^r, \ \ \ 
D_{r,r}^{\sigma,\sigma}(1)=(-1)^r2(\sigma+r), \ \ \
D_{r+1,r}^{\sigma,\sigma}(1)=(-1)^{r}\sqrt{2(r+1)(2\sigma+r)}\ .
\end{equation}

For the sake of convenience, let us define a normalized coupling matrix 
$Y$ from (\ref{y-e}) by
\begin{equation}
Y^{mn}=\frac{y_e^{mn}}{y_e^{00}}=\epsilon'^{m+n}
\sqrt{\frac{(m+n)!}{m!n!}
\frac{\Gamma(2\eta+n)\Gamma(2\lambda+m)\Gamma(2\sigma)}
{\Gamma(2\eta)\Gamma(2\lambda)\Gamma(2\sigma+m+n)}}\ b_{m+n}(\sigma).
\end{equation}
This matrix has a hierarchical structure
\begin{equation}
Y=\left(\begin{array}{lll}
1&\epsilon'
\sqrt{\frac{\eta}{\sigma}}\ b_1(\sigma)&
\epsilon'^2
\sqrt{\frac{(2\eta+1)\eta}{(2\sigma+1)\sigma}}\ b_2(\sigma)\\
\epsilon'
\sqrt{\frac{\lambda}{\sigma}}\ b_1(\sigma)&
\epsilon'^2
\sqrt{\frac{4\eta\lambda}{(2\sigma+1)\sigma}}\ b_2(\sigma)&
\epsilon'^3
\sqrt{\frac{3(2\eta+1)\eta\lambda}{\sigma(\sigma+1)(2\sigma+1)}}\ b_3(\sigma)\\
\epsilon'^2
\sqrt{\frac{(2\lambda+1)\lambda}{(2\sigma+1)\sigma}}\ b_2(\sigma)&
\epsilon'^3
\sqrt{\frac{3(2\lambda+1)\eta\lambda}
{\sigma(\sigma+1)(2\sigma+1)}}\ b_3(\sigma)&
\epsilon'^4
\sqrt{\frac{6(2\eta+1)(2\lambda+1)\eta\lambda}
{\sigma(\sigma+1)(2\sigma+1)(2\sigma+3)}}\ b_4(\sigma)
\end{array}
\right).
\label{Y}
\end{equation}
When the eigenvalues $Y_0,Y_1$,$Y_2$ of this type of matrix
are expanded in the power series of $\epsilon^{\prime 2}$ by
\begin{equation}
Y_0=w_0+O(\epsilon'^2),\hspace{1em}
Y_1=\epsilon'^2w_1+O(\epsilon'^4),\hspace{1em}
Y_2=\epsilon'^4w_2+O(\epsilon'^6),
\label{eq:Y_mn_eigenvalues}
\end{equation}
the leading terms have the expressions
\begin{eqnarray}
w_0&=&Y^{00} , \\
\epsilon'^2w_1&=&\frac{Y^{00}Y^{11}-Y^{01}Y^{10}}{Y^{00}} , \\
\epsilon'^4w_2&=&\frac{
Y^{22}(Y^{00}Y^{11}-Y^{01}Y^{10})
-Y^{21}(Y^{00}Y^{12}-Y^{10}Y^{02})
-Y^{20}(Y^{11}Y^{02}-Y^{01}Y^{12})}
{Y^{00}Y^{11}-Y^{01}Y^{10}} .\nonumber\\
\end{eqnarray}
The straightforward calculation gives the expressions
of $Y_0,Y_1,Y_2$ in the form
\begin{eqnarray}
&&Y_0=1+O(\epsilon'^2), \label{eq:Y_0}\\
&&Y_1=\epsilon'^2
\sqrt{\frac{\eta\lambda}{(2\sigma+1)\sigma}}\ X(\sigma)
+O(\epsilon'^4),\label{eq:Y_1}\\
&&Y_2=\epsilon'^4
\sqrt{\frac{6\eta\lambda(2\eta+1)(2\lambda+1)}
{\sigma(\sigma+1)(2\sigma+1)(2\sigma+3)}}
\ \frac{Z(\sigma)}{X(\sigma)}
+O(\epsilon'^6) ,
\end{eqnarray}
where
\begin{eqnarray}
X(\sigma)&=&2b_2(\sigma)-b_1(\sigma)^2\sqrt{\frac{2\sigma+1}{\sigma}}
 ,\label{X(sigma)}\\
Z(\sigma)&=&2b_2(\sigma)b_4(\sigma)+2b_1(\sigma)b_2(\sigma)b_3(\sigma)
\sqrt{\frac{2\sigma+3}{2\sigma}}\nonumber\\
&&-3b_3(\sigma)^2\sqrt{\frac{2\sigma+3}{6(\sigma+1)}}
-2b_2(\sigma)^3\sqrt{\frac{(\sigma+1)(2\sigma+3)}{6\sigma(2\sigma+1)}}
-b_1(\sigma)^2b_4(\sigma)\sqrt{\frac{2\sigma+1}{\sigma}}.\nonumber\\
\label{Z(sigma)}
\end{eqnarray}

The eigenvalues $Y_0,Y_1,Y_2$ are related to the 
charged lepton masses $m_\tau$,$m_\mu$,$m_e$ through the ratios
\begin{equation}
R_e^0\equiv\left|\frac{Y_1Y_2}{Y_0^2}\right|
=\frac{m_\mu m_e}{m_\tau^2},\hspace{2em}
R_e^1\equiv\left|\frac{Y_0Y_2}{Y_1^2}\right|=\frac{m_\tau m_e}{m_\mu^2}.
\end{equation}
The ratio $R_e^0$ represents the magnitude of the mass hierarchy.
The structure of the mass hierarchy is 
characterized by the ratio $R_e^1$. 
They are given by 
\begin{eqnarray}
R_e^0&=&|\epsilon'|^6\frac{\eta\lambda}{\sigma(2\sigma+1)}
\sqrt{\frac{6(2\eta+1)(2\lambda+1)}{(\sigma+1)(2\sigma+3)}}\ \left|
Z(\sigma)\right|
+O(\epsilon'^8), \label{eq:R_e_0}\\
R_e^1&=&\sqrt{\frac{6(2\eta+1)(2\lambda+1)\sigma(2\sigma+1)}
{\eta\lambda(\sigma+1)(2\sigma+3)}}\
\left|\frac{Z(\sigma)}{X(\sigma)^3}\right|+O(\epsilon'^2) . \label{eq:R_e_1}
\end{eqnarray}
These ratios should be compared with the observed values \cite{ref:PDG}
\begin{eqnarray}
R_e^0({\rm obs})=\frac{m_\mu m_e}{m_\tau^2}
\simeq 1.6\times 10^{-5},\hspace{1em}
R_e^1({\rm obs})=\frac{m_\tau m_e}{m_\mu^2}
\simeq 0.082.
\label{R-obs}
\end{eqnarray}

The remarkable feature of the $SU(1,1)$ model is that,
although $Y_1$ and $Y_2$ are already suppressed
 by $\epsilon^{\prime 2}$
and $\epsilon^{\prime 4}$, respectively, when $\epsilon'<1$,
the factors from the C-G coefficients
realize the unexpectedly large hierarchy
through the strong cancellations within the terms in 
(\ref{X(sigma)}) and (\ref{Z(sigma)})
owing to the fact that each element of $Y$ in (\ref{Y})
is systematically controlled
by the $SU(1,1)$ symmetry.

As an example, let us show this fact
in the simplest case where $S_0=0$ and $S_1=1$.
In this case, $b_n(\sigma)$ takes a form
\begin{equation}
b_n(\sigma)=\sqrt{\frac{\Gamma(2\sigma)}{2^nn!\Gamma(2\sigma+n)}} .
\end{equation}
This gives
\begin{eqnarray}
&&Y_1=-\epsilon'^2\frac{1}{2^2\sigma^2(2\sigma+1)}
\sqrt{\eta\lambda}+O(\epsilon'^4),\\
&&Y_2=\epsilon'^4\frac{1}{2^4\sigma(\sigma+1)^2(2\sigma+1)^2(2\sigma+3)}
\sqrt{\eta\lambda(2\eta+1)(2\lambda+1)}+O(\epsilon'^6) ,
\end{eqnarray}
and then
\begin{eqnarray}
&&R_e^0=
|\epsilon'|^6\frac{1}{2^6\sigma^3(\sigma+1)^2(2\sigma+1)^3(2\sigma+3)}
\eta\lambda\sqrt{(2\eta+1)(2\lambda+1)}+O(\epsilon'^8), \\
&&R_e^1=
\frac{\sigma^3}{(\sigma+1)^2(2\sigma+3)}
\sqrt{\frac{(2\eta+1)(2\lambda+1)}{\eta\lambda}}+O(\epsilon'^2).
\end{eqnarray}
If we simply set $2\eta=2\lambda=\sigma=1$, we have
\begin{equation}
R_e^0=|\epsilon'|^6\frac{1}{2^9\cdot 3^3\cdot 5}
=|\epsilon'|^6\cdot 1.4\times 10^{-5}+O(\epsilon'^8) ,
\hspace{2em}R_e^1=0.2+O(\epsilon'^2) .
\end{equation}
Although the result for $R_e^1$ is unacceptable,
this suggests that we need not introduce unnaturally small $\epsilon'$.
If we chose the weights as
$2\eta=2\lambda=\sigma=1/2$, we obtain the reasonable values
\begin{equation}
R_e^0=|\epsilon'|^6\cdot 1.63\times 10^{-4}+O(\epsilon'^8) ,
\hspace{2em}
R_e^1= 0.083+O(\epsilon'^2).
\label{R^0,1}
\end{equation}
Fitting the observed mass hierarchy 
(\ref{R-obs}) by the leading term of (\ref{R^0,1}),
we obtain $\epsilon'^2= 0.46$.
When we diagonalize the matrix $Y$ numerically with the input
values $2\eta=2\lambda=\sigma=1/2,\ \epsilon'^2=0.46$,
we find
\begin{equation}
R_e^0=1.1\times 10^{-5} ,\hspace{2em}
R_e^1=0.092 .
\end{equation}
This means that the contribution from higher order terms is
subdominant even when $|\epsilon'|\simeq 1$
and does not alter the value from the leading term by the factor beyond $O(1)$.

We may have an expectation, from this rough analysis, that
this minimal $SU(1,1)$ model 
may give a hopeful scheme to understand the characteristic
structure of the mass hierarchy of quarks and leptons.
We will see,
however, this scheme encounters a serious difficulty.

The down-type higgs doublet $h'$ couples not only to charged 
leptons but also to down-type quarks
through $SU(1,1)$ invariant couplings $y_E\bar ELH'+y_D\bar DQH'$. 
Therefore, 
the quantities related to the latter are obtained 
from those of the former by simply replacing the $SU(1,1)$ weights
$\eta$ and $\lambda$ by $\alpha$ and $\gamma$.
Thus, from (\ref{eq:R_e_0}) and (\ref{eq:R_e_1}), 
we obtain the expressions of the  mass ratios for down-type quarks;
\begin{eqnarray}
R_d^0&=&\frac{m_s m_d}{m_b^2}=
|\epsilon'|^6\frac{\alpha\gamma}{\sigma(2\sigma+1)}
\sqrt{\frac{6(2\alpha+1)(2\gamma+1)}{(\sigma+1)(2\sigma+3)}}
\left|Z(\sigma)\right|
+O(\epsilon'^8), \label{eq:R_d_0}\\
R_d^1&=&\frac{m_b m_d}{m_s^2}=
\sqrt{\frac{6(2\alpha+1)(2\gamma+1)\sigma(2\sigma+1)}
{\alpha\gamma(\sigma+1)(2\sigma+3)}}
\left|\frac{Z(\sigma)}{X(\sigma)^3}\right|+O(\epsilon'^2) .\label{eq:R_d_1}
\end{eqnarray}
We show the expressions 
(\ref{eq:R_e_0}), (\ref{eq:R_e_1}), (\ref{eq:R_d_0}), (\ref{eq:R_d_1})
are incompatible with the
observation, when $O(\epsilon'^2)$ corrections are subdominant. 
To see this clearly, 
let us define the cross ratios ${\cal R}$ and ${\cal S}$ by
\begin{equation}
{\cal R}\equiv\bigg(\frac{m_s m_\tau}{m_b m_\mu}\bigg)^2,\hspace{2em}
{\cal S}\equiv\bigg(\frac{m_d m_\mu}{m_s m_e}\bigg)^2.
\end{equation}
Neglecting $O(\epsilon'^2)$ corrections, we find, for any functional
form of $b_i(\sigma)$,
\begin{equation}
{\cal R}= \left(\frac{R_d^0R_e^1}{R_e^0R_d^1}\right)^{2/3}
=\frac{\alpha\gamma}{\eta\lambda} ,
\hspace{2em}
{\cal S}= \left(\frac{R_e^0}{R_d^0}{\cal R}\right)^{-2}
=\frac{(2\alpha+1)(2\gamma+1)}{(2\eta+1)(2\lambda+1)} .
\end{equation}
The restriction of the weights (\ref{weight-restriction})
combines these two expressions to
\begin{equation}
4({\cal R}-{\cal S})\eta\lambda= ({\cal S}-1)(2\sigma+1) ,
\end{equation}
which states, from the positivity of the weights, that
${\cal R}$ and ${\cal S}$ must satisfy
\begin{equation}
1<{\cal S}<{\cal R} \hspace{1em}{\rm or}\hspace{1em} {\cal R}<{\cal S}<1 .
\label{eq:R_S_relation}
\end{equation}
This is the prediction of the minimal model.
The observed values \cite{ref:PDG}
\begin{eqnarray}
{\cal R}({\rm obs})=\bigg(\frac{m_sm_\tau}{m_bm_\mu}\bigg)^2
\simeq 0.1 ,\hspace{2em}
{\cal S}({\rm obs})=\bigg(\frac{m_dm_\mu}{m_sm_c}\bigg)^2
\simeq 100 
\end{eqnarray}
strongly conflict with the above prediction.
The origin of this conflict is the observed values \cite{ref:PDG}
\begin{equation}
R_d^0({\rm obs})=\frac{m_sm_d}{m_b^2}\simeq 1\times10^{-5} ,\hspace{2em}
R_d^1({\rm obs})=\frac{m_bm_d}{m_s^2}\simeq 3 ,
\label{R_d}
\end{equation}
which show, although the magnitudes of the hierarchy $R^0({\rm obs})$'s
are in the same order, 
the structures of hierarchy $R_e^1(\rm obs)\simeq 0.08$ and 
$R_d^1({\rm obs})\simeq 3$ are
too different to compromise by the allowed values of the weights
and the possible higher order corrections of $\epsilon'^2$.
If we try to fit (\ref{R_d}), we need to introduce the higher 
value of the highest weight $S$ for $\Psi^{\prime 0}$ and 
$\Psi^{\prime 1}$, for example $S=3$.\cite{ref:su11-2} 
But in this case, we cannot fit $R_e^0(\rm obs)$ and $R_e^1(\rm obs)$.

One may imagine the difficulty is due to the simplest assumption
(\ref{eq:quark_lepton_no_hierarchy}),  that identifies 
the first three components of the $SU(1,1)$ multiplets with
the three chiral generations.
In fact, we have a strong motivation,\cite{ref:su11-2}
as will be explained at the end of section 4, 
to replace the couplings $Q\bar Q\Psi^F$, etc. in (\ref{W0}) by
\begin{equation}
(Q\bar Q+\bar UU+\bar DD+L\bar L+\bar EE)(\Psi^F+\Psi^{\prime F})
\end{equation}
with VEV 
$\langle\Psi^{\prime F}\rangle=\langle\psi^{\prime F}_0\rangle$ 
and the highest weight $S^{\prime F}=S^F\equiv S$.
It is straightforward to confirm the mixing coefficients take a form
\begin{equation}
U^\ell_{nj}=U^\ell_n\sum_{s=0}^\infty\delta_{j,n+3s}
(-\epsilon^F)^s b^\ell_{ns}(\eta), \hspace{1.5em}n=0,1,2
\end{equation}
with
\begin{equation}
\epsilon^F=\frac{\langle\psi^{\prime F}_0\rangle}
{\langle\psi^{F}_{-3}\rangle},
\hspace{2em}
b^\ell_{ns}(\eta)=\prod_{r=0}^{s-1}
\frac{D^{\eta,\eta}_{3r+n,3r+n}(S)}{D^{\eta,\eta}_{3r+n,3r+n+3}(S)}.
\label{epsilon^F}
\end{equation}
The expression (\ref{y-general}) then gives $y_e^{mn}$ in the form
\begin{equation}
y_e^{mn}=y_EU^e_mU^\ell_n\sum_{r,s=0}^\infty 
(-\epsilon^F)^{r+s}
C^E_{m+3r,n+3s}
b^e_{mr}(\lambda)b^\ell_{ns}(\eta)U'_{m+n+3(r+s)} .
\end{equation}
This clearly shows that the mixing effect $(r+s>0)$ is 
almost negligible.

\section{Extension of the higgs sector}\label{sec:extend_higgs}

The result of the previous section
claims that we should
change the basic setup of the model. 
It is now obvious that the difficulty is 
coming from the assumption that only one higgs multiplet
$H'$ supplies masses to both of charged leptons and down-type quarks.
Thus, the legitimate remedy will be the extension of the higgs sector.
We examine the minimal extension
by doubling the higgs sector 
$a$ $la$ Georgi-Jarlskog.\cite{ref:G-J_relation}

We introduce, in addition to $H'$ and $\bar{H}'$, 
another set of higgs multiplets $K'$ and
$\bar{K}'$ whose weights may not necessarily be equal to 
those of $H'$ and $\bar{H}'$;
\begin{equation}
H'_{-\sigma},\hspace{1em}\bar{H}_\sigma',\hspace{1em}
K'_{-\sigma-\Delta},\hspace{1em}\bar{K}_{\sigma+\Delta}' :
\hspace{2em}\sigma+\Delta> 0 .
\end{equation}
Since we wish to realize MSSM at low energy,
we need a mixing scheme that
generates only one chiral higgs doublet $h'$ 
as a linear combination of the components of 
$H'$ and $K'$ in the way
\begin{equation}
h'_{-\sigma-n}=h'U_n'+[\mbox{massive modes} ],\hspace{1em}
k'_{-\sigma-\Delta-n}=h' V'_n+[\mbox{massive modes} ].
\label{eq:higgs_massless_mode}
\end{equation}
For this purpose, we introduce finite-dimensional
non-unitary $SU(1,1)$ multiplets $\Psi',\Phi',X',\Omega'$, and couple them 
to higgs multiplets by
\begin{equation}
W_{h'}=H'\bar H'\Psi' + K'\bar K'\Phi' + 
K'\bar H'X' + H'\bar K'\Omega' .
\label{Wh'}
\end{equation}
In order to keep the vector-like nature of the model,
$X'$ and $\Omega'$ must be assigned to the same representation
of $SU(1,1)$.
We already know the $SU(1,1)$ invariance requires the weights
of $\Psi'$ and $\Phi'$ to be integer.
On the other hand, that of $X'$ and $\Omega'$ 
is allowed to be integer (Type-I)
or half-integer (Type-II).

Corresponding to the type of $X'$ and $\Omega'$,
the Yukawa couplings of $H'$ and $K'$ to leptons and quarks,
which replace $\bar D Q H'+\bar ELH'$ in (\ref{W1}),
take different forms.
The type-I case, which restricts $\Delta$ to be integer,
allows $H'$ and $K'$ to couple to both
of them by
\begin{equation}
{\mbox{Type-I}}\ :\hspace{2em}
\bar E_\lambda L_\eta (H'_{-\sigma}+K'_{-\sigma-\Delta})+
\bar D_\gamma Q_\alpha(H'_{-\sigma}+K'_{-\sigma-\Delta}) .
\label{Type-I}
\end{equation}
We assume $H'$ takes the lowest coupling.
Thus, we have 
\begin{equation}
\lambda+\eta=\gamma+\alpha=\sigma ,\hspace{2em}
\Delta=0,1,2,3,\cdots .
\end{equation}
On the other hand, the type-II case in (\ref{Wh'}) requires 
$\Delta$ to be half-integer. 
This forbids the simultaneous coupling of 
$H'$ and $K'$ like (\ref{Type-I}), 
and the Yukawa couplings are 
separated by
\begin{equation}
\hspace{-9.5em}\ {\mbox{Type-II}}\ :\hspace{2em}
\bar E_\lambda L_\eta H'_{-\sigma}+
\bar D_\gamma Q_\alpha K'_{-\sigma-\Delta} .
\label{Type-II}
\end{equation}
In this case, both of $H'$ and $K'$ must take the lowest coupling.
Thus, we have 
\begin{equation}
\lambda+\eta=\sigma ,\hspace{2em}\gamma+\alpha=\sigma+\Delta ,\hspace{2em}
\Delta=\Delta_{\rm min},\cdots,-\frac{1}{2},+\frac{1}{2},+\frac{3}{2},
\cdots ,
\label{weight-II}
\end{equation}
where $-\sigma<\Delta_{\rm min}\le -\sigma+1$.

Now, let us proceed to the indispensable 
ingredient of the scheme,
that is, how to generate one chiral higgs $h'$ from doubled higgs sector
through the superpotential (\ref{Wh'}).
It will be natural to expect
each finite-dimensional multiplet acquires non-vanishing VEV
at its single component.
So, we first arbitrarily assume the weights $P,Q,M,$ and $N$ of the VEVs by
\begin{equation}
\langle\Psi'\rangle
=\langle\psi'_P\rangle ,
\hspace{1em}
\langle\Phi'\rangle
=\langle\phi'_Q\rangle ,
\hspace{1em}
\langle X'\rangle
=\langle \chi'_{M}\rangle ,
\hspace{1em}
\langle\Omega'\rangle=\langle\omega'_N\rangle .
\end{equation}
Substituting the expressions (\ref{eq:higgs_massless_mode}) to
the Higgs mass operators (\ref{Wh'}), we have
\begin{eqnarray}
&&H^\prime\bar H^\prime\langle\Psi^\prime\rangle
+K^\prime\bar K^\prime\langle\Phi^\prime\rangle
+K^\prime\bar H^\prime\langle X^\prime\rangle
+H^\prime\bar K^\prime\langle \Omega^\prime\rangle
\nonumber\\
&&=\sum_{n=0}^\infty h^\prime\left(
C_n^{(P)}\langle\psi_P^\prime\rangle
U^\prime_{n+P}
+F^{(M)}_n\langle\chi_{M}^\prime\rangle 
V^\prime_{n+M-\Delta}
\right)\bar h^\prime_{\sigma+n}\nonumber\\
&&\ \ \ \ +\sum_{n=0}^\infty h^\prime\left(
D_n^{(Q)}\langle\phi_Q^\prime\rangle V^\prime_{n+Q}
+F^{\prime(N)}_n\langle\omega_{N}^\prime\rangle 
U^\prime_{n+N+\Delta}
\right)\bar k^\prime
_{\sigma+\Delta+n}+[\mbox{massive modes}]~.\nonumber\\
\label{HK-mass}
\end{eqnarray}
All C-G coefficients 
are given in terms of the general formula (\ref{C-G-2}) by
\begin{eqnarray}
&&C^{(P)}_n=D^{\sigma,\ \sigma}_{n+P,\ n}(S_\Psi) ,\hspace{2.7em} 
D^{(Q)}_n=D^{\sigma+\Delta,\ \sigma+\Delta}_{n+Q,\ n}(S_\Phi) ,
\label{CG-higgs1}\\ 
&&F_n^{(M)}=D^{\sigma+\Delta,\ \sigma}_{n+M-\Delta,\ n}(S_X) ,
\hspace{1em} 
F^{\prime(N)}_n=D^{\sigma,\ \sigma+\Delta}_{n+N+\Delta,\ n}(S_\Omega) ,
\hspace{1.5em}S_\Omega=S_X .
\label{CG-higgs2}
\end{eqnarray}
The disappearance of $h'$ from (\ref{HK-mass}) requires 
the coefficients of $\bar h'_{\sigma+n}$ and $\bar k'_{\sigma+\Delta+n}$
to vanish together;
\begin{eqnarray}
&&C_n^{(P)}\langle\psi_P^\prime\rangle
U^\prime_{n+P}
+F^{(M)}_n\langle\chi_{M}^\prime\rangle 
V^\prime_{n+M-\Delta}=0 ,
\label{rec-1}\\
&&D_n^{(Q)}\langle\phi_Q^\prime\rangle V^\prime_{n+Q}
+F^{\prime(N)}_n\langle\omega_{N}^\prime\rangle 
U^\prime_{n+N+\Delta}=0 .
\label{rec-2}
\end{eqnarray}
If we wish to have only one chiral $h'$,
these recursion equations must determine all elements of
$U'_n$ and $V'_n$ uniquely
by the single input value $U'_0$ or $V'_0$.
This is possible only when 
just one among the four numbers 
$P$, $Q$, $M-\Delta$, and $N+\Delta$ appearing in the subscripts 
of $U'_i$ and $V'_i$ 
is 1 and others are 0.

Therefore, we have 
four cases 
that realize single chiral $h'$;
\begin{eqnarray}
&&\mbox{case A} :\hspace{1em}
P=1 ,\hspace{1em} Q=0 ,\hspace{1em} M=\Delta ,
\hspace{2.5em}N=-\Delta ,\label{A}\\
&&\mbox{case B} :\hspace{1em}
P=0 ,\hspace{1em} Q=1 ,\hspace{1em} M=\Delta ,
\hspace{2.5em}N=-\Delta ,\label{B}\\
&&\mbox{case C} :\hspace{1em}
P=0 ,\hspace{1em} Q=0 ,\hspace{1em} M=\Delta+1 ,
\hspace{0.7em} N=-\Delta ,\label{C}\\
&&\mbox{case D} :\hspace{1em}
P=0 ,\hspace{1em} Q=0 ,\hspace{1em} M=\Delta ,
\hspace{2.5em}N=1-\Delta .\label{D}
\end{eqnarray}
The mixing coefficients $U'_n$ and $V'_n$ 
then satisfy the recursion equations
\begin{eqnarray}
&&\mbox{case A} :\ \ 
U'_{n+1}=\epsilon'\frac{F_n^{(\Delta)}F_{n}^{\prime (-\Delta)}}
{C_n^{(1)}D_{n}^{(0)}}
\ U'_n ,\hspace{2em}
V'_{n+1}=\epsilon'\frac{F_n^{(\Delta)}F_{n+1}^{\prime (-\Delta)}}
{C_n^{(1)}D_{n+1}^{(0)}}V'_n ,\label{A-rec}\\
&&\mbox{case B} :\ \ 
U'_{n+1}=\epsilon'\frac{F_{n+1}^{(\Delta)}F_{n}^{\prime (-\Delta)}}
{C_{n+1}^{(0)}D_{n}^{(1)}}
\ U'_n ,\hspace{2em}
V'_{n+1}=\epsilon'\frac{F_n^{(\Delta)}F_{n}^{\prime (-\Delta)}}
{C_n^{(0)}D_{n}^{(1)}}V'_n ,\label{B-rec}\\
&&\mbox{case C} :\ \ 
U'_{n+1}=\epsilon'\frac{C_n^{(0)}D_{n+1}^{(0)}}
{F_n^{(\Delta+1)}F_{n+1}^{\prime (-\Delta)}}\ U'_n ,\hspace{1.1em}
V'_{n+1}=\epsilon'\frac{C_{n}^{(0)}D_{n}^{(0)}}
{F_{n}^{(\Delta+1)}F_{n}^{\prime (-\Delta)}}\ V'_n ,\label{C-rec}\\
&&\mbox{case D} :\ \ 
U'_{n+1}=\epsilon'\frac{C_n^{(0)}D_{n}^{(0)}}
{F_n^{(\Delta)}F_{n}^{\prime (1-\Delta)}}\ U'_n ,\hspace{1.6em}
V'_{n+1}=\epsilon'\frac{C_{n+1}^{(0)}D_{n}^{(0)}}
{F_{n+1}^{(\Delta)}F_{n}^{\prime (1-\Delta)}}\ V'_n ,\label{D-rec}
\end{eqnarray}
where $\epsilon'$ is given for each case by
\begin{equation}
\mbox{case A, B} :\ \ 
\epsilon'\equiv\frac{\langle\chi'_M\rangle\langle\omega'_N\rangle}
{\langle\psi'_P\rangle\langle\phi'_Q\rangle} ,
\hspace{2.5em}
\mbox{case C, D} :\ \ 
\epsilon'\equiv\frac{\langle\psi'_P\rangle\langle\phi'_Q\rangle}
{\langle\chi'_M\rangle\langle\omega'_N\rangle} ,
\label{epsilon-prime}
\end{equation}
and input values $U'_0$ and $V'_0$ are connected by
(\ref{rec-2}) for case A and C and by (\ref{rec-1}) for case B and D.
We take a phase convention of $h'$ so that $U'_0$ is real and positive.

One may suspect that the couplings (\ref{Wh'})
generate, in addition to $h'$, chiral $\bar h'$ in 
$\bar H'$ and $\bar K'$ in the form
\begin{equation}
\bar h'_{\sigma+n}=\bar h'\bar U_n'+[{\rm massive~modes} ],
~~
\bar k'_{\sigma+\Delta+n}=\bar h' \bar V'_n+
[{\rm massive~modes} ].
\end{equation}
It is straightforward to confirm that the set of VEVs of all
four cases (\ref{A})$\sim$(\ref{D}) gives $\bar U'_n=\bar V'_n=0$. 

The pattern of the VEVs (\ref{A})$\sim$(\ref{D}) is necessary 
for an appearance of single $h'$.
But this is not sufficient.
The realization of $h'$ further requires the normalizable conditions
$\sum_{n=0}^\infty |U'_n|^2<\infty$ and $\sum_{n=0}^\infty |V'_n|^2<\infty$. 
If these conditions were not satisfied, 
$h'$ were an illusion without any physical reality.
That is, we need to have
\begin{equation}
\lim_{n\rightarrow\infty}\left|\frac{U'_{n+1}}{U'_n}\right|<1 ,\hspace{2em}
\lim_{n\rightarrow\infty}\left|\frac{V'_{n+1}}{V'_n}\right|<1 .
\end{equation}
This requires the knowledge of the asymptotic behavior
of $D_{i,j}^{\sigma,\zeta}(S)$ in the limit $i,j\rightarrow\infty$
with $|i-j|$ fixed.
The expression (\ref{C-G-2}) gives
\begin{equation}
D_{i,j}^{\sigma,\zeta}(S)\simeq(-1)^jj^S\frac{(2S)!}
{(S-\bar\Delta)!(S+\bar\Delta)!
\sqrt{(S-\bar\Delta-i+j)!(S+\bar\Delta+i-j)!}} .
\end{equation}
From this asymptotic behavior, we find
\begin{eqnarray}
&&\mbox{case A, B} :\ \ 
\lim_{n\rightarrow\infty}\left|\frac{U'_{n+1}}{U'_n}\right|
=\lim_{n\rightarrow\infty}\left|\frac{V'_{n+1}}{V'_n}\right|
=\frac{|\epsilon'|}{L}n^{2S_X-S_\Psi-S_\Phi} ,
\label{AB-limit}\\
&&\mbox{case C, D} :\ \ 
\lim_{n\rightarrow\infty}\left|\frac{U'_{n+1}}{U'_n}\right|
=\lim_{n\rightarrow\infty}\left|\frac{V'_{n+1}}{V'_n}\right|
=\frac{|\epsilon'|}{L}n^{S_\Psi+S_\Phi-2S_X} ,
\label{CD-limit}
\end{eqnarray}
with case-dependent but $n$-independent number $L$.
This result shows that chiral $h'$ appears for any finite value of 
$\epsilon'$ when the highest weights of the multiplets 
$\Psi'$, $\Phi'$, $X'$, and $\Omega'$ satisfy
\begin{equation}
\mbox{case A, B} :\ \ S_\Psi+S_\Phi>2S_X ,\hspace{2.5em}
\mbox{case C, D} :\ \ 2S_X>S_\Psi+S_\Phi .
\label{non-marginal}
\end{equation}
In these cases, the hierarchy between $U'_n$ and $U'_{n+1}$ 
(and also between $V'_n$ and $V'_{n+1}$) becomes larger
with the increase of $n$.

The situation that faithfully realizes the basic ansatz
$U'_n\propto\epsilon^{\prime n},\ V'_n\propto\epsilon^{\prime n}$ is
a marginal case;
\begin{equation}
S_\Psi+S_\Phi=2S_X .
\label{marginal}
\end{equation}
In this case, $\epsilon'$ must be subject to the constraint
\begin{equation}
|\epsilon'|<L .
\end{equation}

There is a convincing reason to prefer the marginal
assignment (\ref{marginal}).\cite{ref:su11-2}
It is sensible to expect there are plenty of particles in Nature
more than those of MSSM.
Some of them may be related to $H'$ and $K'$ by some symmetry $G$
to form irreducible representations of $G$.
As an illustration, let us consider the grand unified theory (GUT) 
with $G=SU(5)$.\cite{ref:SU(5)GUT,ref:SUSY-SU(5)GUT}
In this case, we inevitably have color-triplet GUT partners
$H'_c$ and $K'_c$ that form $\bf 5$'s of $SU(5)$ with $H'$ and $K'$.
The non-marginal assignment (\ref{non-marginal}) generates, in addition to 
$h'$, color-triplet chiral $h'_c$ in the low energy.
The marginal assignment
solves this notorious
doublet-triplet mass problem in a natural way.
Suppose $X'$ and $\Omega'$ belong to $\bf 1$, and $\Psi'$ and $\Phi'$
to $\bf 24$ of $SU(5)$ in case A.
The difference of the $SU(5)$ C-G coefficients of $\bf 24$ for $H'$
and $H'_c$ ($1:-2/3$) brings
the difference of $L$ for $h'$ and $L_c$ for $h'_c$
by $L_c=(-2/3)^2L$.
This means that, when $\epsilon'$ takes a value so that
$\frac{4}{9}L<|\epsilon'|<L$, 
$h'_c$ becomes an illusion and disappears from physical world
but $h'$ survives
as a chiral superfield.

\section{Details of the Yukawa couplings}

Now, let us discuss the detailed structure of the Yukawa couplings
in Type-I and Type-II schemes.
Since we have no clue at hand on the dynamics to which the non-unitary
finite-dimensional multiplets should be subject,
it will be fair to treat all four patterns of VEVs
(\ref{A})$\sim$(\ref{D}) as the possible candidates.
We omit the mixing effect of quarks and leptons that has been shown
to be negligible.

\subsection*{Type-I scheme}

The Yukawa coupling of $h^\prime$ to leptons in the Type-I scheme
(\ref{Type-I}) is now
\begin{eqnarray}
&&y_E\bar ELH^\prime+y_E^\Delta\bar ELK^\prime\nonumber\\
&&\hspace{1.5em}=\sum_{i,j=0}^\infty 
\left(y_EC^E_{i,j}\bar e_{\lambda+i}l_{\eta+j}h^\prime_{-\sigma-i-j}+
y_E^\Delta\sum_{k=0}^\infty 
C^{E\Delta}_{i,j}\bar e_{\lambda+i}l_{\eta+j}
k^\prime_{-\sigma-k-\Delta}
\delta_{k,i+j-\Delta}\right)\nonumber\\
&&\hspace{1.5em}\longrightarrow \sum_{m,n=0}^2(y_EC^E_{m,n}U^\prime_{m+n}
+y_E^\Delta C^{E\Delta}_{m,n}V^\prime_{m+n-\Delta})\bar e_ml_nh^\prime ,
\end{eqnarray}
that is,
\begin{equation}
y_e^{mn}=y_EC^E_{m,n}U^\prime_{m+n}
+y_E^\Delta C^{E\Delta}_{m,n}V^\prime_{m+n-\Delta} .
\label{new-Yukawa}
\end{equation}
The C-G coefficient $C^E_{m,n}$ 
and $C^{E\Delta}_{m,n}$ are expressed in terms of 
$C_{i,j}^{\lambda,\eta}(\Delta)$
given in (\ref{C-G-1}) by
\begin{equation}
C_{m,n}^E=C_{m,n}^{\lambda,\eta}(0) ,\hspace{1.5em}
C_{m,n}^{E\Delta}=C_{m,n}^{\lambda,\eta}(\Delta) .
\label{C-G-E}
\end{equation} 
Note that $V^\prime_{n<0}=0$.
Therefore, the second term in (\ref{new-Yukawa}) contributes
only to the matrix elements with $m+n\ge \Delta$.
Because $U'_n$ and $V'_n$ are definitely related by (\ref{rec-1})
and (\ref{rec-2}), the coupling matrix $y_e^{mn}$ 
in the Type-I scheme is generally represented in the form
\begin{equation}
y_e^{mn}=y_EC^E_{m,n}\left(1-r_E\theta_{m+n,\Delta}
\frac{C^{E\Delta}_{m,n}}{C^E_{m,n}}\Pi_{m+n}(\sigma)\right)U'_{m+n} ,
\label{yukawa-e-I}
\end{equation}
where $\theta_{m+n,\Delta}=1$ for $m+m\geq\Delta$ and 0
 for $m+n<\Delta$.
To derive the expressions for the numerical factor $r_E$ and the function
$\Pi_{m+n}(\sigma)$, first perform $\Delta$ steps of recursion
from $V'_{m-\Delta}$ to $V'_m$ by 
(\ref{A-rec})$\sim$(\ref{D-rec}),
and rewrite $V'_m$ in terms of $U'_m$ by (\ref{rec-2}) for case A and C
and by (\ref{rec-1}) for case B and D.
The result is
\begin{eqnarray}
&&\mbox{case A}:\ \ r_E=\frac{y_E^\Delta}{y_E}
\frac{\langle\omega'_{-\Delta}\rangle}{\langle\phi'_0\rangle}
(\epsilon')^{-\Delta} ,
\hspace{2em}\Pi_m(\sigma)=
\frac{F^{\prime (-\Delta)}_{m}}{D^{(0)}_{m}}
\prod_{r=1}^{\Delta}\frac{C^{(1)}_{m-r}D^{(0)}_{m-r+1}}
{F^{(\Delta)}_{m-r}F^{\prime (-\Delta)}_{m-r+1}} ,\label{r-A}\\
&&\mbox{case B}:\ \ r_E=\frac{y_E^\Delta}{y_E}
\frac{\langle\psi'_0\rangle}{\langle\chi'_\Delta\rangle}
(\epsilon')^{-\Delta} ,
\hspace{2.7em}\Pi_m(\sigma)=
\frac{C^{(0)}_{m}}{F^{(\Delta)}_{m}}
\prod_{r=1}^{\Delta}\frac{C^{(0)}_{m-r}D^{(1)}_{m-r}}
{F^{(\Delta)}_{m-r}F^{\prime (-\Delta)}_{m-r}} ,\\
&&\mbox{case C}:\ \ r_E=\frac{y_E^\Delta}{y_E}
\frac{\langle\omega'_{-\Delta}\rangle}{\langle\phi'_0\rangle}
(\epsilon')^{-\Delta} ,
\hspace{2.2em}\Pi_m(\sigma)=
\frac{F^{\prime (-\Delta)}_{m}}{D^{(0)}_{m}}
\prod_{r=1}^{\Delta}\frac{F^{(\Delta+1)}_{m-r}F^{\prime (-\Delta)}_{m-r}}
{C^{(0)}_{m-r}D^{(0)}_{m-r}} ,\\
&&\mbox{case D}:\ \ r_E=\frac{y_E^\Delta}{y_E}
\frac{\langle\psi'_{0}\rangle}{\langle\chi'_\Delta\rangle}
(\epsilon')^{-\Delta} ,
\hspace{2.8em}\Pi_m(\sigma)=
\frac{C^{(0)}_{m}}{F^{(\Delta)}_{m}}
\prod_{r=1}^{\Delta}\frac{F^{(\Delta)}_{m-r+1}F^{\prime (1-\Delta)}_{m-r}}
{C^{(0)}_{m-r+1}D^{(0)}_{m-r}} .\label{r-D}
\end{eqnarray}
The expression of $\epsilon'$ is given in
(\ref{epsilon-prime}) for each case.
To preserve the hierarchy 
$y_e^{mn}\propto \epsilon'^{m+n}$ in the coupling matrix
(\ref{yukawa-e-I}), $r_E$ must be $O(1)$.
This requires us to assign the orders of the VEVs to
\begin{equation}
\mbox{case A, C} :\ \ 
\langle\omega'_{-\Delta}\rangle\simeq 
\epsilon^{\prime \Delta}\langle\phi'_0\rangle ,
~~~
\mbox{case B, D} :\ \ 
\langle\psi'_0\rangle\simeq 
\epsilon^{\prime \Delta}\langle\chi'_\Delta\rangle .
\end{equation}

The coupling matrix $y_d^{mn}$ for down-type quarks is obtained from 
(\ref{yukawa-e-I}) by the adequate replacement;
\begin{equation}
y_d^{mn}=y_DC^D_{m,n}\left(1-r_D\theta_{m+n,\Delta}
\frac{C^{D\Delta}_{m,n}}{C^D_{m,n}}\Pi_{m+n}(\sigma)\right)U'_{m+n} ,
\label{yukawa-d-I}
\end{equation}
where
\begin{equation}
C_{m,n}^D=C_{m,n}^{\gamma,\alpha}(0) ,\hspace{1.5em}
C_{m,n}^{D\Delta}=C_{m,n}^{\gamma,\alpha}(\Delta) ,\hspace{1.5em}
r_D=r_E\frac{y_E y_D^\Delta}{y_D y_E^\Delta} .
\label{C-G-D}
\end{equation}

If we are modest and 
hesitate to introduce the multiple hierarchies among the VEVs,
we may be led to take $\Delta=0$ or $\Delta=1$.
When $\Delta=0$, 
we recognize case B and case C are equivalent to
case A and case D, respectively, under the replacement of $\Psi'$
and $\Phi'$, because $H'$ and $K'$ now belong to the same representation.
In these cases, the coupling matrix $y_e^{mn}$ takes a simple form
\begin{eqnarray}
&&\hspace{-6em}\mbox{Type-I-A}_{\Delta=0} :\hspace{2em} 
y_e^{mn}=y_EC^E_{m,n}\left(1-r_E
\frac{F^{\prime(0)}_{m+n}}{D^{(0)}_{m+n}}\right)U'_{m+n} ,
\label{E-yukawa-I-A0}
\\
&&\hspace{-6em}\mbox{Type-I-D}_{\Delta=0} :\hspace{2em} 
y_e^{mn}=y_EC^E_{m,n}\left(1-r_E
\frac{C^{(0)}_{m+n}}{F^{(0)}_{m+n}}\right)U'_{m+n} .
\label{E-yukawa-I-D0}
\end{eqnarray}
When $\Delta=1$, we see the VEVs of case B requires
the multiple hierarchies among the VEVs;
$\langle\omega'_{-1}\rangle\simeq\epsilon^{\prime 2}
\langle\phi'_1\rangle$, 
$\langle\psi'_0\rangle\simeq
\epsilon^{\prime}\langle\chi'_1\rangle$.
The situation is the same for case C.
The VEVs of case A and case D give
\begin{eqnarray}
&&\mbox{Type-I-A}_{\Delta=1} :\hspace{2em} 
y_e^{mn}=y^{\ }_EC_{m,n}^E\left(
1-r_E\theta_{m+n,1}\frac{C_{m,n}^{E\Delta}C^{(1)}_{m+n-1}}
{C_{m,n}^E F^{(1)}_{m+n-1}}\right)U'_{m+n} ,
\label{E-yukawa-I-A1}
\\
&&\mbox{Type-I-D}_{\Delta=1} :\hspace{2em} 
y_e^{mn}=y^{\ }_EC_{m,n}^E\left(
1-r_E\theta_{m+n,1}\frac{C_{m,n}^{E\Delta}F^{\prime (0)}_{m+n-1}}
{C_{m,n}^E D^{(0)}_{m+n-1}}\right)U'_{m+n} .
\label{E-yukawa-I-D1}
\end{eqnarray}

\subsection*{Type-II scheme}

The Yukawa coupling matrices of the Type-II scheme 
have a simple form
\begin{equation}
y_e^{mn}=y^{\ }_EC^E_{m,n}U'_{m+n} ,\hspace{2em}
y_d^{mn}=y^{\ }_DC^D_{m,n}V'_{m+n} ,
\end{equation}
with C-G coefficients given in (\ref{C-G-E}) and (\ref{C-G-D}).
The weights are now restricted by (\ref{weight-II}).
The natural hierarchy of the coupling matrices 
is realized by the recursion equations
(\ref{A-rec})$\sim$(\ref{D-rec}).
The input values $U'_0$ and $V'_0$ are connected by the definite
relation
\begin{equation}
\mbox{case A, C}:\ \ 
V'_0=-\frac{\langle\omega'_{-\Delta}\rangle}{\langle\phi'_0\rangle}
\frac{F^{\prime(-\Delta)}_0}{D_0^{(0)}}\ U'_0 ,
~~
\mbox{case B, D}:\ \ 
V'_0=-\frac{\langle\psi'_{0}\rangle}{\langle\chi'_\Delta\rangle}
\frac{C^{(0)}_0}{F_0^{(\Delta)}}\ U'_0 .
\label{Type-II-input}
\end{equation}

\subsection*{Characteristics of the coupling matrix}

Let us first point out one important fact
that concerns both Type-I and Type-II schemes.
The expression of $\epsilon'$, which is concisely represented in 
(\ref{epsilon-prime}), 
shows that $\epsilon'$ transforms under the $U(1)_H$ 
subgroup of $SU(1,1)$
as if it is a ``VEV with weight $-1$'' in all cases A$\sim$D,
that precisely coincides with (\ref{epsilon}) of the minimal model.
This is not an accident but a consequence of the requirement 
that only one down-type
higgs doublet $h'$ is 
realized as a chiral superfield.
This implies that what has been meant in terms of the vague
phrase ``natural hierarchy'' is not a smallness of $\epsilon'$
but its definite transformation property under $U(1)_H$.
\cite{ref:Frog-Nielsen,ref:anomalous-U(1)}
The magnitude of the hierarchy is a consequence of the non-Abelian 
group structure of $SU(1,1)$.

There are several points which should be mentioned on the property
of the Type-I scheme.
First of all,
this scheme admits two extra free parameters 
$r_E$ and $r_D$. They are in general complex numbers, but 
we realize, from their relation (\ref{C-G-D}), they have common phase
because all of the coupling constants $y_E$, $y_E^\Delta$, 
$y_D$, $y_D^\Delta$ are real numbers under the ``fundamental principle''
of P-C-T-invariance.
This phase should be a physical one independent of the phase
convention of the VEVs.
This is confirmed from its
expressions (\ref{r-A})$\sim$(\ref{r-D}),
that insure $r_E$ to have ``weight 0'' and do not move under 
$U(1)_H$ rotation.
Since we have no explicitly
complex number in the basic framework,
it will be natural to expect
\begin{equation}
\frac{r_E}{|r_E|}=\pm\frac{r_D}{|r_D|}=e^{i\pi{q}/{p}}
\end{equation}
with some set of integers $p$ and $q$.
The simplest candidate is of course $q/p=[\mbox{integer}]$, that is, 
$r_E$ and $r_D$ are pure real (Option-1).
This phase assignment brings a fascinating chance for us to have a 
texture-zero \cite{ref:texture-0,ref:texture-0-1,ref:texture-0-2}
in $y_e^{mn}$ and $y_d^{mn}$
at specific values
of $r_E$ and $r_D$ 
by the cancellation of two terms in the coupling matrix.
This texture-zero appears in a remarkable pattern in $\Delta=0$
scheme.
From their expressions (\ref{E-yukawa-I-A0}) and (\ref{E-yukawa-I-D0}), 
we find 
all of the matrix elements with common $m+n$ vanish together.
It should, however, be sensible to expect this cancellation
occurs, at most, approximately rather than  exactly, 
because we have no principle in the framework that guarantees
the exact cancellation.
The uncomfortable aspect of the Option-1 phase assignment is
that it causes the instability of the natural hierarchy of 
coupling matrices at various values of $r_E$ and $r_D$
through the accidental cancellation
in the eigenvalues of the coupling matrix.
That is, would-be $O(1)$ eigenvalue becomes $O(\epsilon^{\prime 2})$
and would-be $O(\epsilon^{\prime 2})$ eigenvalue becomes
$O(\epsilon^{\prime 4})$.
Such a phenomenon does not occur in the minimal $SU(1,1)$ model when 
the marginal assignment $S_0=S_1\equiv S$ is taken in (\ref{H'mass});
the condition $X(\sigma)=0$ in (\ref{eq:Y_1}) leads to inaccessible 
$\sigma$-independent constraint $S^2+S=0$.
The optimum that maximally stabilizes the hierarchy in Type-I scheme
will be
$q/p=[\mbox{half-interger}]$, that is, $r_E$ and $r_D$ are pure
imaginary (Option-2).

The second point is on the phase of $\epsilon'$ which is 
also complex in general.
However, this phase is dependent on the
phase convention of the VEVs because it has ``weight $-1$''.
Therefore we can always rotate out the phase of $\epsilon'$
so that it becomes real and positive.
This does not mean at all that this phase is 
completely unphysical.
When we have another quantity $\epsilon$ which coherently moves
with $\epsilon'$
under $U(1)_H$ rotation, the relative phase of $\epsilon'$ 
and $\epsilon$ is a physical observable.
This $\epsilon$ surely exists in our framework 
as a mixing parameter in the up-type higgs sector,
which we are not discussing in this paper.
This relative phase and the phase of $r_E$ and $r_D$
are reflected on the phases in the CKM and MNS matrices.

The third point on the Type-I scheme, 
which may be the most appealing point,
 is that this scheme allows us to assign all quarks 
and leptons
to the common $SU(1,1)$ representation.
When this assignment is adopted in 
$\Delta=[\mbox{odd number}]$ schemes,
the first part in the coupling matrices
(\ref{yukawa-e-I}) and (\ref{yukawa-d-I})
becomes a symmetric matrix but the second part
which contains $r$'s becomes antisymmetric
owing to the symmetry relation 
$C_{i,j}^{\lambda,\eta}(\Delta)=(-1)^\Delta C_{j,i}^{\eta,\lambda}(\Delta)$
of (\ref{C-G-1}).
This means that the Option-2 phase assignment realizes
the hermitian Yukawa coupling matrices  \cite{ref:texture-0-2} 
 under the real $\epsilon'$
phase convention.
On the other hand, 
$\Delta=[\mbox{even number}]$ 
schemes give perfectly symmetric coupling matrices.

The last point which should be mentioned 
on the Type-I scheme is the assumption in (\ref{Type-I}).
In principle, there is a possibility that $K'$ takes the lowest
coupling, for example, in $\bar DQK'$.
In this case, $\bar DQH'$ is forbidden and $y_d^{mn}$ takes simpler
form without parameter $r_D$.
When both of $\bar ELK'$ and $\bar DQK'$ take the lowest coupling,
the situation reduces to the minimal model, that has been ruled out. 

The typical property of the Type-II scheme is that,
under the real $\epsilon'$ phase convention,
$y_e^{mn}$ and $y_d^{mn}$ become pure real matrices 
up to overall phases connected by (\ref{Type-II-input}).
This means that, when the up-type higgs doublet $h$ also takes the Type-II
mixing scheme (or Type-I scheme with Option-1 phase assignment), 
the origin of the phase of the CKM matrix is
solely the relative phase of $\epsilon$ and $\epsilon'$,
$\delta\equiv\arg(\epsilon/{\epsilon'})$.
This restricts the form of the CKM matrix to
\begin{equation}
V_{\rm CKM}=O_u^TPO_d ,\hspace{2em}
P\equiv\left(\begin{array}{ccc}e^{2i\delta}&0&0\\0&e^{i\delta}&0\\
0&0&1\end{array}\right),
\end{equation}
where $O_u$ and $O_d$ are the real orthogonal matrices, 
standing on the right when 
the ``reversed'' $(m,n=2,1,0)$ coupling matrices
$y_u^{mn}$ and $y_d^{mn}$ with real and positive 
$\epsilon$ and $\epsilon'$
are diagonalized, respectively.
The analysis of Type-II scheme will be an exciting subject 
because we have no free parameter like $r_E$
and $r_D$.
All mass ratios must be reproduced in terms of $\epsilon'$ and the 
weights of each multiplet, that we failed in the minimal 
$SU(1,1)$ model.

\section{Summary and discussions}

The simplest non-Abelian noncompact group $SU(1,1)$ gives us
an excellent viewpoint on what is realized in the low energy physics
when it is introduced, in the framework of the vector-like theory,
 as a symmetry group of the generations of quarks and leptons.
It gives, in terms of its spontaneous breaking, an answer
 to the questions why three generations of quarks and
leptons are simply repeated and why they have, never-the-less, rich
hierarchical mass structures. 

In this paper we first 
investigated the structure of the Yukawa coupling
matrices $y_e^{mn}$ and $y_d^{mn}$
of the minimal $SU(1,1)$ model, which contains single
down-type higgs multiplet $H'$,
and found this minimal model gives a definite prediction 
on the mass ratios, that is  phenomenologically far 
unacceptable.
If Nature really uses $SU(1,1)$, this result means
that the higgs sector should have richer structure than that of the 
minimal model.
Following this observation, we formulated the minimal
extension of the higgs sector by incorporating additional higgs
multiplet $K'$. 

The extension of the higgs sector is a non-trivial subject because
the success of MSSM insists that there is 
 only one down-type higgs doublet $h'$
at low energy.
This requires the special mixing scheme that realizes single $h'$
at low energy from doubled higgs sector which consists of $H'$ and $K'$
(and their conjugates).
We found throughout the formulation  
that the general framework of the model
admits this mixing scheme in quite restrictive way.
As a result, it was shown that the possible forms of the 
Yukawa couplings are classified to
two types, Type-I and Type-II, each of which has four patterns
of mixing, A, B, C, and D.
Each scheme reveals its own property in the specific structure
of the coupling matrices.

Although the principle is simple, the resulting expressions 
of the coupling matrices are much complicated.
It is difficult, at present, to make definite statement on 
which type of scheme is
phenomenologically most preferable.
The decision should wait the result of the global analysis 
of the full scheme of the model 
that contains the mixing scheme also for the
up-type higgs $h$.
It should be mentioned, however, that the preliminary numerical analysis
on the Type-I scheme
shows there are some sets of 
parameters that reproduce the mass ratios 
reasonably.
In the following, we give the typical examples.
All mass ratios should be compared with the ratios at the GUT-scale.
\cite{ref:mass}
In each case, the marginal assignment (\ref{marginal}) is adopted.

In the analysis, we assumed
a universal quark and lepton assignment
\begin{equation}
2\alpha=2\gamma=2\eta=2\lambda=\sigma 
\end{equation}
within the restricted range of the discrete values 
$\sigma=1/2,1,3/2,2,5/2,3,7/2$.
We found the Type-I-A$_{\Delta=1}$ scheme with $S_{\Psi}=S_{\Phi}=S_X=3$,
which requires $|\epsilon'|<L=2.0_5$, accepts the set of parameters
\begin{equation}
\left\{\begin{array}{ll}\sigma=1,&\epsilon'=0.77\\ r_E=\pm1.33,&r_D=\pm
 0.41
\end{array}\right.
~\mbox{gives}~\left\{\begin{array}{ll}
m_\mu/m_\tau=0.060_0,&m_e/m_\mu=0.0048_8\\ m_s/m_b=0.017_0,&
m_d/m_s=0.044_5\end{array}\right..
\end{equation}
Also in the Type-I-D$_{\Delta=0}$ scheme with 
$S_\Psi=2$, $S_\Phi=0$, $S_X=1$, which requires
$|\epsilon'|<L=1.0_6$, 
\begin{equation}
\left\{\begin{array}{ll}\sigma=3/2,&\epsilon'=0.72\\ r_E=0.94,&r_D=1.38
\end{array}\right.
~\mbox{gives}~\left\{\begin{array}{ll}
m_\mu/m_\tau=0.057_0,&m_e/m_\mu=0.0048_2\\ m_s/m_b=0.020_1,&
m_d/m_s=0.053_1\end{array}\right..
\end{equation}
The parameter search for the Type-II scheme is now under investigation.

We would like to
 conclude this paper with a briefly discussion on the ``full set''
of the superpotential $W$ of the model. 
For definiteness, we assume the Type-I$_{\Delta>0}$ 
mixing scheme for both of up-type and down-type higgs sectors.
$W$ consists of the four parts;
\begin{equation}
W=W_{\rm MSSM}+W_{\rm N}+W_{\rm M}+W(\mbox{finite dim.}).
\end{equation}
The first part $W_{\rm MSSM}$ reproduces MSSM up to $\mu$-term;
\begin{eqnarray}
W_{\rm MSSM}&=&(Q\bar Q+\bar UU+\bar D D+L\bar L+\bar EE)
(\Psi^F+\Psi^{\prime F})\nonumber\\
&&+H\bar H\Psi+K\bar K\Phi+K\bar HX+H\bar K\Omega\nonumber\\
&&+H'\bar H'\Psi'+K'\bar K'\Phi'+K'\bar H'X'+H'\bar K'\Omega'\nonumber\\
&&+\bar UQ(H+K)+\bar DQ(H'+K')+\bar EL(H'+K')
+QQD+Q\bar U\bar L\nonumber\\
&&+U\bar Q(\bar H+\bar K)+D\bar Q(\bar H'+\bar K')
+E\bar L(\bar H'+\bar K')+\bar Q\bar Q\bar D+\bar QUL .
\label{W-MSSM}
\end{eqnarray}
We assume $H$ and $H'$ take the lowest coupling in the operators
in the fourth line.

The second part $W_{\rm N}$ is responsible to the neutrino masses.
Since the Majorana mass operator of a type
$NN\Psi^N$ is forbidden by the $SU(1,1)$ symmetry,
the see-saw mechanism \cite{ref:see-saw} does not work so efficiently
in the present framework.
The adequate alternative is to 
introduce $SU_2$ triplets.\cite{ref:su11-3}
The observed large mixing angles 
\cite{ref:neutrino-experiment-1,ref:neutrino-experiment-2,
ref:neutrino-experiment-3} in the MNS
matrix seem to require some sets of triplets $T^i$ and their conjugates;
\begin{eqnarray}
W_{\rm N}&=&\sum_{i,j}T^i\bar T^j\Psi^{ij}
+LL\sum_i\bar T^i
+(HH+HK+KK)\sum_iT^i\nonumber\\
&&\hspace{6em}+\bar L\bar L\sum_iT^i
+(\bar H\bar H+\bar H\bar K+\bar K\bar K)\sum_i\bar T^i .
\end{eqnarray}
We must be careful so that the VEVs $\langle\Psi^{ij}\rangle$
do not generate massless particles in $T^i$ and $\bar T^i$.
Then, the integration of $T^i$ and $\bar T^i$ produces the neutrino
mass operator
\begin{equation}
\sum_{m,n=0}^2\kappa_\nu^{mn}\ell_m h\ell_n h 
\end{equation}
with the coupling matrix $\kappa_\nu^{mn}\simeq O(M^{-1})$.
The structure of $\kappa_\nu^{mn}$
is sensitive to the assignment 
of the weights of each multiplet
and the pattern of the VEVs $\langle\Psi^{ij}\rangle$.

The third part $W_{\rm M}$ contains $SU_2$ singlets
$R_{(\rho+\sigma)/2}$, $R'_{(\rho+\sigma)/2+1}$, 
$S_{\rho+\sigma}$ and their conjugates.
It is responsible to the $\mu$-term $\mu hh'$;
\begin{eqnarray}
W_{\rm M}&=&
R\bar R'\Psi^R+R'\bar R\Psi^{\prime R}+
R\bar R'\Psi^M+R'\bar R\Psi^{\prime M}+S\bar S\Psi^S \nonumber\\
&&+HH'S+RR\bar S+\bar H\bar H'\bar S+\bar R\bar RS .
\end{eqnarray}
The VEVs  $\langle\psi_{1}^R\rangle$, 
$\langle\psi_{-1}^{\prime R}\rangle$, 
$\langle\psi_0^M\rangle$,
$\langle\psi_0^{\prime M}\rangle$, 
$\langle\psi_0^S\rangle$ 
generate a set of massless $r$ and $\bar r$ in $R$ and 
$\bar R$, respectively,
to which the superheavy $\bar S$ and $S$ couple.
When the supersymmetry breaking terms characterized by 
the scale $m_{\rm {SUSY}}\simeq 10^{2\sim 3}$GeV
are incorporated in the scheme,
$r$ and $\bar r$ acquire the VEVs in the intermediate scale
$\sqrt{Mm_{\rm SUSY}}$.
The integration of $S$ and $\bar S$ eventually induces the $\mu$-term
at the scale $\mu\simeq m_{\rm SUSY}$.
At this time, we must take care in the assignment of the weight of $L$
so that the dangerous couplings $L(H+K)R$ and $L(H+K)\bar R$
are forbidden by the weight constraint.

The resulting mass spectrum in the low energy effective theory
contains, in addition to the MSSM particles, 
several neutral particles coming from $r$ and $\bar r$.
Most of them have masses of order $m_{\rm SUSY}$,
but their couplings to higgses $(hh')$ are suppressed 
by the factor $\sqrt{{m_{\rm SUSY}}/{M}}$.
What is unexpected is the emergence of the long-range force
mediated by the ``exactly massless'' pseudo-scalar
 Nambu-Goldstone (N-G) boson $G^0$,
which originates from the spontaneous breakdown of the Peccei-Quinn
$U(1)_{\rm PQ}$ symmetry.\cite{ref:PQ_symmetry}

The appearance of the $U(1)_{\rm PQ}$ symmetry is a consequence
of the $SU(1,1)$ symmetry, which restricts the possible couplings 
of matter multiplets by the weight constraint.
The finite-dimensional multiplets will not share the $U(1)_{\rm PQ}$
charge since the $SU(1,1)$ symmetry does not impose so stringent
constraint on their couplings in $W(\mbox{\rm finite dim.})$.
The N-G boson $G^0$ couples to quarks and leptons through the mixing
with the pseudo-scalar particle $a^0$ in the MSSM with the mixing fraction
of order $\sqrt{m_{\rm SUSY}/M}$.
At a glance, it seems to be almost the invisible axion.\cite{ref:axion}
The essential difference is now the $U(1)_{\rm PQ}$ symmetry is 
an ``exact'' symmetry free from gauge anomalies because the basic framework
of the model is purely vector-like.
Therefore, $G^0$ does not couple to two photons nor two gluons through
gauge anomalies.

In principle, we cannot deny a possibility that the $U(1)_{\rm PQ}$
symmetry is explicitly broken in the couplings of the matter
multiplets to the superheavy multiplets 
$(Z^d,\bar Z^d)$ which we have been discarding.
In this case, the $U(1)_{\rm PQ}$ symmetry 
revives in the low energy effective theory
as if it is an ``anomalous'' symmetry.
When this is the case, $G^0$ will acquire a mass
suppressed by the huge mass $M$ of the superheavy multiplets.
So, it will be prudent to imagine $G^0$ has a tiny mass 
$m_G\simeq (m_{\rm SUSY})^{n+1}/M^n$ 
with some positive integer $n$.
Not only the $U(1)_{\rm PQ}$ symmetry, but also the baryon number 
$U(1)_B$ 
and the lepton number $U(1)_L$ symmetries,
though explicitly broken, do not suffer
from gauge anomalies.
These results may give significant impacts on the understanding
of the present universe.

Finally, the fact that the $U(1)_{\rm PQ}$ symmetry does not suffer from 
gauge anomalies means that $G^0$ loses the role
as a solution to the strong CP problem.\cite{ref:strong-CP}
As an alternative, 
we have now a exact  P-C-T-invariance \cite{ref:strong-CP-zee}
at the fundamental level.
Therefore we may have a chance to solve the problem 
based on this invariance supplemented by the $U(1)_{\rm PQ}$
symmetry.
We would like to leave this subject to the future study.

To be honest, we should state that we do not have, at present, 
a well established background
that allows us to make such an unconventional  scenario.
We are anticipating the off-shell property of the superstring theory,
when fully clarified, 
gives a reliable support.

\section*{Acknowledgments}

The authors would like to thank K. Yoshioka for 
enlightening suggestion, K. Harada for stimulating encouragement
and A. Watanabe, K. Kojima, H. Sawanaka and X. Tata
for helpful discussions.
This work is supported in part by the Grant-in-aid for the scientific
research on priority area ($\sharp$ 441)
``Progress in elementary particle physics of the 21st century through 
discoveries of Higgs boson and supersymmetry'' (No.16081209)
from the Ministry
of Education, Science, Sports, and Culture of Japan.

\end{document}